\documentclass[apj]{emulateapj}
\usepackage[dvips]{color}
\usepackage{rotating}

\usepackage[dvips]{color}
\usepackage{apjfonts}
\usepackage{graphicx}
\slugcomment{}

\shorttitle{CN Anomalies in the Halo System}
\shortauthors{Carollo et al.}

\begin{document}

%% LaTeX will automatically break titles if they run longer than
%% one line. However, you may use \\ to force a line break if
%% you desire.

\title{CN Anomalies in the Halo System and the Origin of Globular Clusters in the Milky Way}

\author{Daniela Carollo\altaffilmark{1}}
\affil{Dept. of Physics and Astronomy - Astronomy, Astrophysics and Astrophotonic Research Center\\ Macquarie University - North Ryde, 2109, NSW, Australia}
\email{daniela.carollo@mq.edu.au}

\author{Sarah L. Martell}
\affil{Australian Astronomical Observatory, North Ryde, 2109, NSW, Australia}
\email{smartell@aao.gov.au}

\author{Timothy C. Beers\altaffilmark{2}}
\affil{National Optical Astronomy Observatory, Tucson, AZ, 85719, USA}
\email{beers@noao.edu}

\author{Ken C. Freeman}
\affil{Research School of Astronomy \& Astrophysics, Australian National
University\\ \& Mount Stromlo Observatory, Cotter Road, Weston, ACT, 2611, Australia}
\email{kcf@mso.anu.edu.au}

\altaffiltext{1} {INAF-Osservatorio Astronomico di Torino, Pino Torinese, Italy}

\altaffiltext{2} {Department of Physics \& Astronomy  and JINA: Joint Institute for Nuclear Astrophysics,\\
Michigan State University, E. Lansing, MI  48824 USA}

\begin{abstract}

We explore the kinematics and orbital properties of a sample of red
giants in the halo system of the Milky Way that are thought to have
formed in globular clusters, based on their anomalously strong UV/blue
CN bands. The orbital parameters of the CN-strong halo stars are
compared to those of the inner- and outer-halo populations as described
by Carollo et al., and to the orbital parameters of globular clusters
with well-studied Galactic orbits. The CN-strong field stars and the
globular clusters both exhibit kinematics and orbital properties similar
to the inner-halo population, indicating that stripped or destroyed
globular clusters could be a significant source of inner-halo field
stars, and suggesting that both the CN-strong stars and the majority of
globular clusters are primarily associated with this population.

\end{abstract}


\section{Introduction}  %% section 1

Although globular clusters were once held up as the prototype of simple
stellar populations, the presence of multiple stellar populations in globulars
is now well-recognized. Evidence for this complexity is found in both
the elemental abundance distributions of individual cluster stars and
from clearly separable multiple sequences in well-measured
color-magnitude diagrams. The chemical pattern most useful to identify
multiple populations is primarily the abundances of light elements
formed by proton-capture nucleosynthesis in the later stages of stellar
evolution. Light-element abundance inhomogeneities, such as the C-N,
O-Na, and Mg-Al anticorrelations, have been found among stars on the red
giant branches in essentially all the globular clusters where sufficient
data exists (e.g., Gratton et al. 2001; Ramirez \& Cohen 2002; Kayser et
al. 2008; Carretta et al. 2009a, 2009b and references therein; Smolinski
et al. 2011b). During the evolution of globular clusters, a large
fraction of stars may have been lost through early violent relaxation
following gas expulsion, mass loss from the most massive stars
(Baumgardt et al. 2008), and the evaporation of a significant fraction
of stars in two-body encounters over long timescales (McLaughlin \& Fall
2008). First-generation stars born in globular clusters that have
migrated into the halo system cannot be readily distinguished from stars
born outside of the clusters based on their chemical abundances. The
situation differs for second-generation stars, because
their peculiar chemical compositions are believed to be obtained only as
a result of their formation inside the deep gravitational potential well
of a globular cluster; this abundance signature acts as a ``chemical
tag'' that provides the opportunity to identify them even after they
have been lost to the halo field.

Earlier studies of halo-star chemistry (e.g., Pilachowski et al. 1996;
Stephens \& Boesgaard 2002; Gratton et al. 2004; Venn et al. 2004) did
not find any of these second-generation ``migrant'' stars, bolstering
the idea that only star formation in globular clusters is able to
produce the characteristic light-element abundance anomalies. More
recently, second-generation stars\footnote{Hereafter, we refer to all
later generations of stars in globular cluster as ``second-generation,''
even though they may have been born during distinct bursts of star
formation.} have been identified in the halo field by Martell \& Grebel
(2010) and Martell et al. (2011), using medium-resolution spectroscopic
data for giants from the Sloan Extension for Galactic Understanding and
Exploration (SEGUE; Yanny et al. 2009). The CN anomalies were found
among SEGUE giants with low or normal carbon abundances (inferred from
the strength of the CH G-band at 4300 $\hbox{\AA}$) that exhibit
unusually strong absorption in the UV/blue CN band (3883 $\hbox{\AA}$),
from which it is inferred that they possess high nitrogen abundances.
Additional evidence for their presence outside of globular clusters is
provided by Carretta et al. (2010), using a compilation of literature
abundance data, and by Ramirez et al. (2012), using high-resolution
spectra of the nearby Nissen \& Schuster (2010) sample of halo dwarf
stars.

In order to put the CN-anomolous stars found in the halo into their
proper context, it is important to recognize that ideas concerning the
nature of the halo have evolved over the past few years. For example,
Carollo et al. (2007, 2010) have argued that the halo of the Milky Way
comprises at least two smooth stellar components, the inner and outer
halos, possessing different peak metallicities ([Fe/H]$_{inner}$ $\sim
-$1.6; [Fe/H]$_{outer}$ $\sim -$2.2), different spatial distributions,
and different kinematics. The inner halo has a flatter density profile
than the nearly spherical outer halo, and has almost zero mean rotation,
while the outer halo exhibits a significantly retrograde rotation. The
transition from dominance by the inner-halo population to the outer-halo
population occurs in the range 15-20 kpc from the Sun.

In this paper, we explore the kinematics and orbital properties of the
CN-strong field giants from the sample of Martell et al. (2011), as well
as for a subset of globular clusters with available proper motions, in
order to assess whether CN-strong field stars are better associated with
the inner- or outer-halo population, and to infer the likely fate of
their parent globular clusters. This paper is organized as follows. In
Section 2, we discuss the origin of the chemical abundance variations in
globular clusters, and the possible connection with the Galactic halo
system. Section 3 describes the data and the selection of the field-star
sample, together with the derivation of the CN line strengths. In
addition, this section describes the selection of a sample of globular
clusters with available proper motions. In Section 4, we describe the
derivation of the kinematics and orbital parameters for the field star
and globular cluster samples. Section 5 presents an analysis of these
datasets. Our main results are summarized in Section 6, along with a
brief discussion of their implications.

\section{Globular Clusters and the Galactic Halo}  %% section 2

\subsection{The Origin of Chemical Abundance Variations in Globular
Clusters}

The observed light-element abundance variations in globular clusters are
often explained in the context of multiple generations of stars within
the clusters. Stars with atypical light-element abundances were likely
to have formed from material enriched by the ejecta of earlier stellar
generations in the cluster (Gratton et al. 2001, 2004; Ramirez \& Cohen
2002; Carretta et al. 2010). Suggestions for the polluters include
asymptotic giant branch (AGB) stars (Cottrell \& Da Costa 1981;
Parmentier et al. 1999; Ventura et al. 2001), rapidly rotating massive
stars (Decressin et al. 2007), and massive binary stars undergoing mass
transfer (de Mink et al. 2009). First-generation stars in globular
clusters are expected to have been stars with typical Population II
compositions (Truran \& Arnett 1971), similar to that of the field halo
stars. Typically, one-third to two-thirds of the stars in a given
globular cluster are thought to be second- (or later-) generation
objects, and exhibit the distinct light-element anomalies (e.g., Kraft
1994; Carretta et al. 2009a).

The relatively high observed ratios of second- to first-generation stars
in globular clusters creates a significant problem for
multiple-generation models of globular cluster evolution -- it is not
possible for the present-day first-generation stars to have produced
sufficient material to pollute the present-day second generation. A
top-heavy IMF for the first generation has been suggested as a possible
solution to this ``mass budget problem'' (e.g., Cannon et al. 1998), but
most current globular cluster formation models (e.g., D'Ercole et al.
2008, 2010; Conroy 2012) assume that the first generation of stars was
initially much more massive (by a factor of 10 to 20) than it is today.
This additional mass at an early point in cluster evolution results in
more sources for stellar nucleosynthesis feedback, and also raises the
escape velocity, making it it more likely for clusters to retain
sufficient polluted gas to form a second generation of stars. This
immediately implies that many or most of the first-generation stars that
initially formed in globular clusters have been subsequently lost to the
halo field populations.

Theoretical models predict that as much as 90-95\% of the
first-generation stars have been lost from globular clusters at
relatively early times (D'Ercole et al. 2008; Vesperini et al. 2010;
Conroy 2012; Vesperini et al. 2012). It is not yet clear if the
second-generation stars were lost all at once, or during later multiple
episodes of globular cluster tidal disruption. Martell et al. (2011)
adopt this scenario to predict that $\sim$17\% of the halo field stars
(exhibiting both first- and second-generation abundance patterns) were
born in globular clusters, while Schaerer \& Charbonnel (2011) estimate
that 5-8\% of halo field stars originally formed in globular clusters.

So far, we have considered the chemical evolution of globular clusters
in a two-generation scenario, in which both generations of star
formation went on within the cluster itself. The very young and massive
LMC cluster NGC 2070 suggests a possible variation on this scenario. NGC
2070 is a globular-like star cluster with a mass of about 5 $\times$
10$^{5}$ M$_{\odot}$ (Bosch et al. 2009), and an age of only two million
years (Massey \& Hunter 1998). The cluster is immersed in the massive 30
Dor nebula, which contains about 4 $\times$ 10$^{6}$ M$_{\odot}$ of HI
(S. Kim, private communication), HII (Kennicutt 1984), and CO (Pineda
et al. 2009), plus about 4 $\times$ 10$^{4}$ M$_{\odot}$ of hot gas
(Wang 1999). Active star formation is going on around the NGC 2070
cluster; some of the stars have masses $>$ 100 M$_{\odot}$ (Massey \&
Hunter 1998), and some of the surrounding young stars are older than
the cluster itself. This spatially extended and ongoing burst of star
formation started before the formation of the cluster, and has
pressurized the environment and contributed to its chemical evolution.

Although the metallicity of the 30 Dor region is higher than that of the
Galactic halo clusters, the 30 Dor system may be much like the early
globular-cluster-forming fragments envisaged by Searle \& Zinn (1978).
If this kind of environment is typical of the formation of the halo
globular clusters, then the star formation in an extended region around
the cluster may provide the first generation of stars, and the cluster
itself is the second generation.

The first generation of stars forms from the background interstellar
medium, and is loosely bound to the cluster. Eventually most of them
will escape into the halo. The second generation (the cluster stars)
forms partly from the background ISM and partly from infalling gas that
has been further enriched by the evolution of the massive stars of the
surrounding first generation. Although these second-generation stars are
initially bound to the cluster, some will escape into the halo during
the dynamical evolution of the cluster, and may be recognized as the
Martell et al. halo stars with CN anomalies.

In this scenario, the mass of the first generation need not be tightly
related to the mass of the cluster itself, so the ratio of
first-generation to second-generation stars escaping into the halo is
likely to vary from cluster to cluster. In particular, the mass of
first-generation stars now within the cluster is not required to be
sufficiently large to produce the fusion-processed material leading to
the abundance offsets seen in the second generation. Furthermore, the
ongoing star formation in the region surrounding the cluster will also
produce stars with CN anomalies, as its star formation continues and it
evolves chemically. These CN-enhanced stars would escape and contribute
to the halo's population of anomalous stars. The low total number of
such stars observed in the halo puts a limit on the total number of
CN-enhanced stars that have come in to the halo, either as escapees from
the clusters or from their surrounding regions.

Which (if either) of these two enrichment scenarios pertains remains
uncertain. A comparison of the properties of the CN-strong stars in the
clusters and in the Galactic halo system may provide a useful guide.

\subsection{Possible Connections with the Halo System of the Galaxy}

Carollo et al. (2010) demonstrated that the flattened inner-halo
population is essentially non-rotating, with V$_{\phi}$ = 7 $\pm$ 4
km~s$^{-1}$, while the near-spherical outer-halo population exhibits a
significant retrograde signature, with V$_{\phi}$ $\sim -$80 km~s$^{-1}$
(where V$_{\phi}$ is the Galactocentric rotational velocity). The
velocity ellipsoids of these populations differ as well, such that
($\sigma_{V_{R}}$, $\sigma_{V_{\phi}}$, $\sigma_{V_{Z}})$ = (150, 95,
85) km~s$^{-1}$ for the inner halo and (159, 165, 116) km~s$^{-1}$ for
the outer halo, evaluated in a Galactocentric cylindrical reference
frame. Kinman et al. (2012) presented similar results, based on samples
of RR Lyrae stars chosen without kinematic bias (a transition from a
flattened, essentially non-rotating inner halo to a retrograde spherical
outer halo beyond about 12.5 kpc). Hattori et al. (2013) have used blue
horizonal-branch (BHB) stars with available metallicities and radial
velocities from the Sloan Digital Sky Survey (SDSS; Gunn et al. 2006; York et al. 2000) in
order to demonstrate that the mean rotational velocity of the very
metal-poor ([Fe/H] $< -2.0$) BHB stars significantly lags behind that of
the relatively more metal-rich ([Fe/H] $> -2.0$) BHB stars. Futhermore,
the relatively more metal-rich BHB stars are dominated by stars with
eccentric orbits, while the very metal-poor BHB stars are dominated by
stars on rounder, lower-eccentricity orbits. Both of these results are
consistent with with dual halo described by Carollo et al. (2007, 2010).

Carollo et al. (2012) have used the dual halo paradigm to account for
the well-known observed increase of the frequency of carbon-enhanced
metal-poor (CEMP) stars with decreasing metallity (see Beers \&
Christlieb 2005, and references therein), as well as for the increase of
CEMP frequency with distance from the Galactic plane (Frebel et al.
2006). Beers et al. (2012) offers additional lines of evidence for the
existence of the dual halo. Most recently, An et al. (2013) have used
photometric estimates of stellar metallicity for stars in SDSS Stripe
82, along with available proper motions, to argue that, even in the
relatively nearby volume (5-8 kpc from the Sun), the observed
metallicity distribution function (MDF) (coupled with the kinematics) of
the halo is incompatible with a single population of stars. Chemical,
kinematic, and spatial signatures for a dual halo have also been
recently found in high-resolution numerical simulations of Milky
Way-like galaxies incorporating baryons (e.g., Zolotov et a. 2010; Font
et al. 2011; McCarthy et al. 2012, Tissera et al. 2012).

Tissera et al. (2013) point out that an important distinction should be
made between the \emph{Inner Halo Population} (IHP) and the
\emph{Inner Halo Region} (IHR), as well as between the \emph{Outer Halo
Population} (OHP) and the \emph{Outer Halo Region} (OHR). Based on the
results of Carollo et al., the IHP of the Milky Way possesses an MDF
peaked at [Fe/H] $\sim -$1.6, extending towards both higher and
lower metallicities, including significant numbers of stars at very low
metallicity, [Fe/H] $< -$2.0. The IHR of the MW is located between 5 and
15-20 kpc, where the IHP is the dominant contributor of stars in the
metallicity range $-$2.0 $<$ [Fe/H] $< -$1.0. Due to the strong
metallicity segregation between the inner and outer halos, most of the
stars in the IHR with metallicity below $\sim$ $-$2.0 belong to the OHP.
The OHP of the Milky Way has an MDF peaked at [Fe/H] $\sim -$2.2,
extending towards both higher and lower metallcities, including
stars with [Fe/H] $> -$2.0. The OHR of the Milky Way is located beyond
$\sim$ 20 kpc, where the OHP dominates in the low-metallicity regime,
[Fe/H] $< -$2.0. The majority of the stars at higher metallicity and
located in the OHR likely belong to the overlapping IHP, or are members
of bound substructures, such as streams, which are not members of the
diffuse stellar component. In this context, the distinction between
inner- and outer-halo objects (stars or globular clusters), based solely
on their Galactocentric distance or metallicity, has to be reconsidered.
An object at Galactocentric distance beyond 20 kpc is located in the
OHR, but it well be a member of the IHP.

\section{Selection of the Samples of CN-strong Field Stars and Globular
Clusters}

The Martell \& Grebel (2010) and Martell et al. (2011) studies of halo
field giants drew their data from the SDSS-II/SEGUE-1 (Abazajian et al.
2009; Yanny et al. 2009) and SDSS-III/SEGUE-2 (Aihara et al. 2011;
Eisenstein et al. 2011; Rockosi et al, in preparation) surveys,
respectively. Both SEGUE surveys were spectroscopic extensions of SDSS,
with the goal of acquiring broad wavelength-coverage,
moderate-resolution ($R \simeq 2000$) optical spectra of stars in
specific Galactic populations. A few examples of those populations are G
and K disk dwarfs (e.g., Lee et al. 2011a; Cheng et al. 2012;
Schlesinger et al. 2012), white dwarf-main sequence binaries
(Rebassa-Mansergas et al. 2012), and distant halo BHB stars (e.g., Xue
et al. 2011). To make the spectra readily useful to the broader
community, the SEGUE Stellar Parameter Pipeline (SSPP) was developed to
estimate metallicities, effective temperatures, surface gravities, and
radial velocities for all stars observed as part of SDSS/SEGUE. The SSPP
uses a variety of methods, including photometric calibrations, template
matching, and spectral indices; details can be found in Lee et al.
(2008a; 2008b), Allende Prieto et al. (2008), Lee et al. (2011b), and
Smolinski et al. (2011a).

The sample selection for the Martell \& Grebel (2010) and Martell et al.
(2011) studies is described thoroughly in those papers, and briefly
summarized here. A generous initial selection was made based on
SSPP-derived parameters: ${\rm log(g)} \leq 3.0$, ${\rm [Fe/H]} \leq
-1.0$, ${\rm (g-r)_{0}} \geq 0.2$, $\sigma_{\rm log(g)}
\leq 0.5$, $\sigma_{\rm [Fe/H]} \leq 0.5$, and a mean signal-to-noise
ratio per pixel of 20 or greater. That initial set was then reduced to
include only likely red giant branch stars by dividing it into 0.2-dex
bins in [Fe/H], and rejecting all stars further than 3-sigma in (g-r)
$_{0}$ color from the mean red giant branch color-magnitude sequence in
that bin. The signal-to-noise requirement was augmented to ensure
high-quality data in the blue spectral features central to their
investigation, requiring that the mean signal-to-noise per pixel in the
wavelength range $4000 \leq \lambda \leq 4100$ be at least $15$.
Carbon-enhanced stars were removed from the sample based on the ${\rm
C}_{\rm 2}$ indices defined in Martell \& Grebel (2010). Stars with
${\rm [Fe/H]} \leq -1.8$ were also removed, because the the CN and CH
bands become quite weak at low metallicity (see, e.g., Shetrone et al.
2010 and Smolinski et al. 2011b for examples of the limits on
band-strength analysis in low-metallicity globular clusters). We have
selected those stars from the Martell et al. (2011) sample that have
available proper motions, which means that the star satisfies additional
criteria designed to eliminate spurious reported motions (see Munn et
al. 2004).\footnote{Note that all proper motions have been corrected for
the systematic error described by Munn et al. (2008).}. Also, stars
belonging to the SDSS/SEGUE fields that fall in the direction of the
Sagittarius stream were removed, in order to excise possible
contaminants. After these selections, the remaining number of stars is
N$_{\rm Tot}$ = 1583; there are N$_{\rm CN}$ = 42 among these stars with
strong CN features.

We have also selected a sample of Galactic globular clusters (hereafter,
GCs) with available proper motions from the
literature\footnote{http://www.astro.yale.edu/dana/gc.html}, in order to
compare the properties of these GCs with those of the CN-strong stars,
and discuss them in the context of the inner- and outer-halo
populations. The sample comprises 59 GCs for which positions, absolute
proper motions, distances, and radial velocities are listed. In this
compilation, the locations of the clusters, distances from the Sun,
radial velocities, and metallicities are taken from the Harris (1996) database
(2010 update\footnote{http://www.physics.mcmaster.ca/~harris/mwgc.dat}).
Errors in the distances are taken to be 10\% of the stated distance. The
absolute proper motions are with respect to distant galaxies, QSOs,
and/or millisecond pulsars, and are on the ICRS system with respect to
the Hipparcos system (in some cases Tycho-2), or with respect to a
kinematic model of the Galaxy. See the web page describing this effort
for more details. The average error in the proper motions is $\sim$ 1
mas yr$^{-1}$. In our analysis, we have removed the globular clusters in
the direction of the Sagittarius and Canis Major dwarf galaxies, in
particular: Pal~12 (Irwin 1999; Palma et al. 2002; Bellazzini et al.
2003; Cohen 2004; Carretta et al. 2010; Law \& Majewski 2010), NGC~4147
(Bellazzini et al. 2003; Carretta et al. 2010; Law \& Majeski 2010),
NGC~4590 (Dinescu et al. 1999; Palma et al. 2002; Casetti-Dinescu et al.
2010; Forbes \& Bridges 2010\footnote{In this paper NGC~4590 is argued
to be associated to the Canis Major dwarf galaxy.}; Dalessandro et al.
2012). NGC~5466 (Palma et al. 2002; Bellazzini et al. 2003) is likely
associated with Sagittarius, while NGC~1851, NGC~1904, NGC~2298, and
NGC~2808 are clusters likely associated with Canis Major (Forbes
\& Bridges 2010). With such a selection, the final sample comprises 51
globular clusters (referred to below as GC$_{PM}$). The kinematics of
the remaining GCs for which proper motions are not available are
considered below in the context of a Frenk \& White (1980) analysis.
From the Harris database we have removed the GCs in the direction of
Sagittarius and Canis Major, and those with Galactocentric distance $>$
50 kpc, which may be associated with other dwarf galaxies. This
subsample contains 78 GCs.

%Note that the sample of GCs is complete up to 10-15 kpc.

%{\bf Here add the GC with no proper motions}
\newpage
\section{Analysis}

\subsection{Derivation of Stellar Parameters}

Estimates of T$_{\rm eff}$, log g, and [Fe/H] for the field stars in our
sample were obtained from the most recent version of the SSPP; typical
internal errors are $\sigma_{\rm Teff}$ $\sim$ 125 K, $\sigma_{\rm log
g}$ $\sim$ 0.25 dex, and $\sigma_{\rm [Fe/H]}$ $\sim$ 0.20 dex. The
external errors in these determinations are of similar size. Due to
recent updates of the SSPP, the metallicities of some of the stars in
our present sample differ slightly from those used by Martell et al.
(2011), which were taken from the seventh and eighth SDSS data releases
(DR7, Abazajian et al. 2009; DR8, Aihara et al. 2011), so that the
present data set includes a handful of stars with metallicities outside
the original range of $-1.8 \le$ [Fe/H] $\le -1.0$. The CN and CH band
strengths in Martell \& Grebel (2010) and Martell et al. (2011), which
we have adopted here, were measured using the indices $S(3839)$ and
$S(CH)$, defined in Norris et al. (1981) and Martell et al. (2008),
respectively. These indices measure the magnitude difference between the
integrated flux in a region of spectrum containing the feature of
interest and the integrated flux in a nearby region of spectrum
unaffected by the feature of interest, in the sense that stronger
absorption in the feature produces a larger band strength. The
differential index $\delta S(3839)$ was calculated following the method
of Norris et al. (1981), by fitting a straight line to the CN-normal
stars in the $S(3839)$ versus absolute magnitude plane, and taking the
difference between the measured band strength and that line at fixed
magnitude.

Distances have been adopted using the approach described by Martell et
al. (2011). As described in that paper, heliocentric distances were
calculated by a straightforward photometric parallax method from the
observed SDSS $(g-r)_{0}$ colors (where reddening corrections were
applied from Schlegel et al. 1998), and interpolating
within the 12-Gyr Dartmouth isochrones (Dotter et al. 2008) of
appropriate metallicity to each star's color to find its absolute $r$
magnitude, then converting the resulting $(r-M_{r})_{0}$ distance
modulus into a distance. Monte-Carlo sampling of the errors on $(g-r)
_{0}$ was used to estimate errors on the resulting heliocentric
distances, typically on the order of 10-15\%. Galactocentric distances were
determined geometrically using the IDL routine {\it lbd2xyz}, available
through Goddard Space Flight Center's online IDL Astronomy
Library\footnote{http://idlastro.gsfc.nasa.gov/idllibsrch.html}.

\subsection{Derivation of Space Motions and Orbital Parameters}

Proper motions, used in combination with distance estimates and radial
velocities, provide the information required to calculate the full space
motions (the components of which are referred to as U,V,W) of our
program stars with respect to the Local Standard of Rest (LSR; defined
as a frame in which the mean space motions of the stars in the Solar
Neighborhood average to zero). The velocity component U is taken to be
positive in the direction toward the Galactic anticenter, the V
component is positive in the direction of Galactic rotation, and the W
component is positive toward the North Galactic Pole. Corrections for
the motion of the Sun with respect to the LSR are applied during the
course of the calculation of the full space motions; here we adopt the
values (U,V,W) = ($-9$,12,7) km~s$^{-1}$ (Mihalas \& Binney 1981). For
the purpose of our analysis it is also convenient to obtain the
rotational component of a star's motion about the Galactic center in a
cylindrical frame; this is denoted as V$_{\phi}$, and is calculated
assuming that the LSR is on a circular orbit with a value of 220
km~s$^{-1}$ (Kerr \& Lynden-Bell 1986). It is worth noting that our
assumed values of R$_{\sun}$ (8.5 kpc) and the circular velocity of the
LSR are both consistent with two recent independent determinations of
these quantities by Ghez et al. (2008) and Koposov et al. (2009).
Bovy et al. (2012) have recently determined, on the basis of accurate
radial velocities for stars in the APOGEE sub-survey of SDSS-III, that
the circular velocity of the LSR is close to 220 km~s$^{-1}$.

The orbital parameters of the stars, such as the perigalactic distance
(the closest approach of an orbit to the Galactic center), r$_{peri}$,
and apogalactic distance (the farthest extent of an orbit from the
Galactic center), r$_{apo}$, of each stellar orbit, the orbital
eccentricity, $e$, defined as $e$ = (r$_{apo}$ $-$ r$_{peri}$)
/(r$_{apo}$ + r$_{peri}$), as well as Z$_{max}$ (the maximum distance of
a stellar orbit above or below the Galactic plane), are derived by
adopting an analytic St\"ackel-type gravitational potential (which
consists of a flattened, oblate disk, and a nearly spherical massive
dark-matter halo; see the description given by Chiba \& Beers 2000,
Appendix A). Typical errors on the orbital parameters (at Z$_{max}$ $<$
50 kpc; Carollo et al. 2010) are: $\sigma_{r_{peri}}$ $\sim$ 1 kpc,
$\sigma_{r_{apo}}$ $\sim$ 2 kpc, $\sigma_{ecc}$ $\sim$ 0.1,
$\sigma_{Z_{max}}$ $\sim$ 1 kpc.  These same methods have been applied
to the sample of GCs with available proper motions. Proper motions,
distances, and radial velocities for the stars in the Martell et al.
(2011) subsample are listed in Table 1, while the derived space motions
and orbital paramaters are listed in Table 2. An analysis of the
kinematics of the sample of GCs, including those without available proper
motions, is reported below.

\section{Results}

\subsection{Halo Field Stars}

The left-hand column of panels in Figure 1 shows the index
$\delta$S(3839) for the CN-normal stars (black dots) and CN-strong stars
(red dots), as a function of Galactocentric distance, $r$ (top panel),
and as a function of the vertical distance, ${\rm |z|}$ (bottom panel).
Note that the sample is limited at small distances by the SDSS/SEGUE
bright limit of $g \sim 14$, and at large distances by the requirement
that the typical S/N per pixel in the blue (faint) end of the spectra be
at least 15. These selection effects on the bright and faint ends of the
data set operate equally on CN-strong and CN-normal stars, so that the
ratio of the two is not affected (see also Section 3.3 of Martell et al.
2011). Note that the CN-strong stars are concentrated in the IHR, and
their frequency drops rapidly beyond 20 kpc, as previously pointed out
by Martell et al. (2011). The distribution of CN-normal and CN-strong
stars as a function of the vertical distance ${\rm |z|}$ shows that most
of the CN-strong stars are located in the region 1 kpc $< {\rm |z|} <$ 8
kpc.

The right-hand column of panels in Figure 1 shows the index
$\delta$S(3839) for the CN-normal stars (black dots) and CN-strong stars
(red dots), as a function of the maximum Galactocentric distance
achieved by stars during their orbits, $r_{max}$ (top panel), and as a
function of the maximum vertical distance achieved by stars during their
orbits, $Z_{max}$ (bottom panel). In Carollo et al. (2007), it was noted
that most stars of the IHP do not possess orbits that take them beyond
15-20 kpc. By way of contrast, stars of the OHP can reach distances well
beyond 20 kpc in their orbits.\footnote{Figure 6 of Carollo et al.
(2007), supplemental material.} Similarly, the CN-strong stars in the
top-right panel of Figure 1 exhibit apogalactic distances that are
mostly located between 5 kpc and 20 kpc, and few orbits beyond 20 kpc,
in agreement with the behavior of the stars of the IHP. Another
remarkable feature is that the great majority of the orbits of the
CN-strong stars are located within Z$_{max}$ < 15 kpc, again
corresponding to the IHR, where the IHP dominates in the metallicity
range $-$2.0 $<$ [Fe/H] $< -$1.5. We have verified that the
distributions of CN-normal and CN-strong stars as a function of $r$ or
$|z|$ does not change when the entire dataset of Martell et al. is
considered, including the stars without available proper motions.

Figure 2 shows the derived V$_{R}$,V$_{\phi}$,V$_{Z}$ velocity components
in the Galactocentric cylindrical reference frame, as a function of
metallicity, for the selected samples of CN-normal stars (black dots) and
CN-strong stars (red dots). As can be appreciated from inspection of the
middle panel, the CN-strong stars at higher metallicity ([Fe/H] $> -$1.1)
exhibit highly prograde rotational velocities, consistent with that
expected for members of the thick-disk and metal-weak thick-disk (MWTD)
components, $\langle$V$_{\phi}$$\rangle$ = 185 km s$^{-1}$ and
$\langle$V$_{\phi}$$\rangle$ = 125 km s$^{-1}$, respectively (Carollo et
al. 2010). Even though the Martell et al. (2011) sample has been selected
to belong primarily to the halo field, it is reasonable to expect some
contamination from the thick disk and metal-weak thick disk. However, the
prograde features in the rotational velocity distribution may not be
simply related to these components. Indeed, a more careful examination of
the rotational velocity as a function of $Z_{max}$ reveals that the highly
prograde stars are still present in regions dominated by the halo system,
$Z_{max} >$ 5 kpc. These stars are most likely members of substructures;
we defer a detailed analysis to a future paper.

We have selected a subsample of metal-poor stars, with [Fe/H] $<-$ 1.5,
to reduce possible contamination from the thick disk and metal-weak
thick disk, and Z$_{max}$ < 15 kpc, in order to avoid the substructures
present in the Martell et al. (2011) data. With these cuts in
metallicity and Z$_{max}$, the total number of stars is N$_{Star}$ =
360; there are N$_{CN}$ = 10 CN-strong stars. Figure 3 (top) shows the
Galactocentric rotational velocity distribution of the selected
subsample of stars. The left panel represents the low-metallicity
CN-normal subsample, while the right panel shows the low-metallicity
CN-strong stars. The mean rotational velocity and dispersion for the
CN-normal stars is $\langle$V$_{\phi}$$\rangle$ = $-19 \pm$ 9 km s$^{-1}$,
and $\sigma_{V_{\phi}}$ = 114 $\pm$ 6 km s$^{-1}$, consistent with
membership in the IHP ($\langle$V$_{\phi}$$\rangle$ = 7 $\pm$ 4 km
s$^{-1}$, and $\sigma_{V_{\phi}}$ = 95 $\pm$ 2 km s$^{-1}$; Carollo et
al. 2010).
%Note that the slightly higher prograde velocity found for the
%Martell et al. (2011) low-metallicity CN-normal subsample could be due
%to residual contamination from the substructures, which dominate the
%sample at higher metallicities.
A two-sample Kolmogorov-Smirnoff (K-S)
test of the distributions of rotational velocity for the low-metallicity
CN-normal and CN-strong stars is unable to reject the hypothesis that
they were drawn from the same parent population ($p = 0.34$). Note that
the number of stars with highly retrograde velocities in the
low-metallicity subsample is very small, N$_{retr}$ = 29 at V$_{\phi} <
-$100 km s$^{-1}$, and N$_{retr}$ = 6 at V$_{\phi} < -$200 km s$^{-1}$,
respectively. Among these groups of stars, none of them are CN strong.

Carollo et al. (2007, 2010) have shown that the IHP is dominated by
high-eccentricity orbits, while the OHP exhibits a much more uniform
distribution of eccentricities (see Figure 4 of the supplemental
material in Carollo et al. 2007 and Figure 5 of Carollo et al. 2010). We
have used the eccentricity parameter to better quantify the connection
between the IHP and the CN-strong stars. Figure 3 (bottom panels) shows
the eccentricity distribution for the selected subsamples of stars. As
before, the left panel represents the low-metallicity CN-normal
subsample, while the right panel shows the low-metallicity CN-strong
stars. Inspection of these panels reveals that the CN-normal stars at
low metallicity are dominated by high-eccentricity orbits ($e >$ 0.5),
in agreement with the eccentricity distribution of the IHP. The
CN-strong stars are also dominated by high-eccentricity orbits. A
two-sample K-S test of eccentricity distribution for the low-metallicity
CN-normal and CN-strong stars is unable to reject the hypothesis that
they were drawn from the same parent population ($p = 0.64$).

\subsection{Comparison with Galactic Globular Clusters}

The general properties of the GCs in our sample with available absolute
proper motions are typical of the Milky Way's cluster population, in terms
of their spatial and metallicity distributions. The metal-rich portion of
the sample ([Fe/H] $> -$1.0) is concentrated towards the center of the
Galaxy, lies close to the Galactic plane, and is rapidly rotating. The
metal-poor portion ([Fe/H] $< -$1.5) occupies a more spherically symmetric
region surrounding the Galactic center, and has a slightly prograde mean
rotational velocity. The top-left panel of Figure 4 shows the spatial
distribution projected onto the YZ plane, in a Galactocentric Cartesian
reference system; the red dots denote GCs at low metallicity, [Fe/H] $<
-$1.5. These GCs are mostly concentrated within $|$Z$|$ $\sim$ 15 kpc. The
top-right panel of Figure 4 shows metallicity for the GCs as a function of
Galactocentric distance. As seen in the figure, the GC sample is primarily
located within 0 $< r <$ 15 kpc, and exhibits two metallicity peaks (as
previously shown by Zinn 1985) -- a small one at [Fe/H] $\sim -$0.6 and a
dominant one at [Fe/H] $\sim -$1.6. The marginal histograms in the right
panel show the distribution of $r$ and [Fe/H]. The bottom panels of Figure
4 show the spatial distribution in the YZ plane and the metallicity as a
function of the Galactocentric distance for the sample of 129 GCs selected
from the Harris database, including those with
available proper motions.

The upper row of panels in Figure 5 shows the rotational velocity
distribution (V$_{\phi}$, Galactocentric cylindrical reference frame)
for the entire GC$_{PM}$ sample in the left panel, and for the
metal-poor subsample ([Fe/H] $< -$1.5) in the right panel. In the left
panel, the highly prograde feature (V$_{\phi}$ $\sim$ 150-180 km
s$^{-1}$) in the velocity distribution is associated with the metal-rich
subsample, while the slightly prograde or non-rotating velocity
distribution is associated with the metal-poor subsample. The dot-dashed
curve in the right panel indicates a Gaussian fit to the distribution,
with mean rotational velocity $\langle$V${_{\phi}}$$\rangle$ $\sim$ 1 km
s$^{-1}$ and dispersion $\sigma_{V_{\phi}}$ $\sim$ 136 km s$^{-1}$.
These values are consistent with membership in the IHP, perhaps with
some contamination from a higher-dispersion population. A two-sample K-S
test of the rotational velocity distribution of the low-metallicity GCs
and the CN-normal stars is unable to reject the hypothesis that they
were drawn from the same parent population ($p = 0.5$). A similar null
result is obtained for a two-sample K-S test of the rotational velocity
distribution of the low-metallicity GCs and the CN-strong stars ($p =
0.2$).

The lower row of panels in Figure 5 shows the eccentricity distribution
for the sample of globular clusters with no selection in metallicity
(left panel), and at low metallicity, [Fe/H] $< -$1.5 (right panel). In
the left panel, the low-eccentricity values are associated with the
metal-rich subsample, while the highly eccentric orbits are associated
with the metal-poor subsample. In the right panel, the subsample at low
metallicity is dominated by high-eccentricity orbits, which are typical
of the IHP. A two-sample K-S test applied to the eccentricity
distribution of the metal-poor GCs and the metal-poor CN-normal stars
sample is unable to reject the hypothesis that they were drawn from the
same parent population ($p = 0.9$). The same test applied to the
eccentricity distribution of the metal-poor GCs and the metal poor
CN-strong stars sample is also unable to reject the hypothesis that they
were drawn from the same parent population ($p = 0.9$).

We have performed a kinematic analysis for the sample of 129 GCs
(comprising the 51 objects with available proper motions and the 78 GCs
with no proper motion available) selected from the Harris database, and
culled as described in Section 3.  Our aim is to explore the
rotational properties of a more extended sample of GCs, and to check
for consistency with the results reported above, which were based on the
subset of clusters with available proper motions. We follow the
procedure described by Frenk \& White (1980), which makes use of
distance and observed radial velocities alone (along with assumed
axisymmetry) in order to estimate the rotation and dispersion of
Galactic tracer populations. From this approach, the mean rotational
velocity derived for the subsample of GCs with Galactocentric distance
$r <$ 15 kpc and metallicity [Fe/H] $< -$1.5 (32 GCs) is V$_{rot}$ = 24
$\pm$ 28 km s$^{-1}$, while the dispersion is $\sigma_{los}$ = 99 $\pm$
13 km s$^{-1}$. These results are in agreement with the values obtained
for the subsample of GCs with available proper motions. Similar results
are obtained when the sample at $r <$ 30 kpc and [Fe/H] $< -$1.5 (44
GCs) is considered, V$_{rot}$ = 38 $\pm$ 36 km s$^{-1}$ and
$\sigma_{los}$ = 88 $\pm$ 10 km s$^{-1}$. We conclude that our kinematic
results, based on the subsample of GCs with available proper motions, is
not unduly biased as a result of this selection.

\section{Summary and Discussion}

We have analyzed the sample of red giant stars in the halo fields
selected by Martell et al. (2011) from the SEGUE-1 and SEGUE-2 surveys,
and determined their kinematic and orbital parameters. After removing
possible contamination from one or more substructures, mainly found with
metallicities above [Fe/H] $\sim -1.5$, we have selected a subsample of
stars for which we have examined the rotational velocity distribution
and the orbits. Also, a sample of globular clusters with available
proper motions from the literature has been assembled, and its kinematic
and orbital properties have been compared to those of the field stars.
\\\\

Our main results can be summarized as follows:

\begin{itemize}

\item The CN-strong stars are located primarily at Galactocentric radii
we associate with the IHR,  $r <$ 15-20 kpc, where the IHP dominates in
the metallicity range covered by this sample.

\item The CN-strong stars occupy orbits that primarily populate
Z$_{max} <$ 15 kpc, typical for the orbits of IHP stars.

\item The CN-strong stars exhibit orbits with apogalactic distances
below 20 kpc, in agreement with the IHP stars.

\item The rotational behavior of the low-metallicity subsample of
Martell et al. (2011) ([Fe/H] $< -$1.5) is typical of the IHP, with mean
velocity and dispersion $\langle$V$_{\phi}$$\rangle$ = $-$19 $\pm$ 9 km
s$^{-1}$, and $\sigma_{V_{\phi}}$ = 114 $\pm$ 6 km s$^{-1}$,
respectively.

\item The eccentricity distribution of the low-metallicity subsample of
Martell et al. (2011) ([Fe/H] $< -$1.5) is also typical of the IHP, with
primarily high-eccentricity orbits.

\item The CN-strong stars in the low-metallicity regime exhibit a rotational
velocity distribution consistent with that of the IHP, and which does
not differ from the rotational velocity distribution of the
low-metallicity CN-normal stars.

\item The CN-strong stars in the low-metallicity regime exhibit an eccentricity
distribution consistent with that of the IHP.

\item None of the stars with CN anomalies possess highly retrograde
orbits.

\item The subsample of CN-normal stars at higher metallicity, [Fe/H]
$> -1.5$, exhibit some evidence for membership in substructures, to be
considered in a future paper.

\item The sample of low-metallicity ([Fe/H] $< -1.5$) Galactic globular
clusters with available proper motions exhibits a rotational velocity
distribution, velocity dispersion, and eccentricity distribution
consistent with that of the IHP and the CN-strong stars.

\end{itemize}

\vspace{5 mm}

\subsection{Implications for the Formation of the Halo System
and the Connection with Galactic Globular Clusters}

Modern high-resolution cosmological simulations at high redshift ($z >$
3) suggest that globular clusters formed in the central cores of giant,
high-density clouds of massive sub-Galactic fragments (or primordial
mini-halos) (e.g., Bekki 2012). The host progenitor galaxies
hierarchically merge onto the main body of the parent galaxy, and are
tidally disrupted. However, the globular clusters that they hosted are
sufficiently dense to survive accretion by the main galaxy. In these
simulations, most globular clusters form in sub-halos of mass M $\geq$
10$^{9}$ M$_{\odot}$, with masses proportional to the amount of gas
present, typically, M$_{GC} \sim 10^{5}$ M$_{\odot}$. These simulations
reproduce the distributions of cluster mass, size, and metallicity
consistent with those of the Galactic metal-poor clusters (Kravtsov \&
Gnedin 2005; Prieto \& Gnedin 2008).

Recent high-resolution simulations of Milky Way-like galaxies that
include prescriptions to account for baryonic material are also able to
reproduce the global properties of the inner and outer components of the
Milky Way's stellar halo system. In particular, they match well with the
observed shift of the stellar MDF towards lower values with increasing
Galactocentric distance, and the observed shear in the mean rotational
velocity between components (Zolotov et al. 2010; Font et al. 2011;
McCarthy et al. 2012; Tissera et al. 2012), as described by Carollo et
al. (2007, 2010).

According to the simulations, the IHP is likely to have formed from the
rapid dissipational mergers of a number of relatively massive clumps. Star
formation within these massive clumps (both pre- and post-merger) would
quickly drive up the mean metallicity. The different rotational
and orbital properties of stars in the OHP component of the Milky Way
clearly indicates that the formation of the outer halo is distinct from
that of the inner halo and the disk components, likely through
dissipationless accretion of lower-mass subsystems within a pre-existing
dark matter halo. A more detailed examination of the nature of the
assembly of the IHP and OHP is presented by Tissera et al. (2013).

The fact that the CN-strong stars exhibit spatial distributions,
rotational velocities, and orbital properties in agreement with the IHP
provides important clues on the origin and fate of GCs in the Milky Way.
The primordial sub-Galactic fragments of higher mass and gas content
presented favorable conditions to form GCs in the inner cores of giant
high-density clouds. By way of contrast, smaller-mass fragments may not
have had sufficient masses of gas to form GCs. These lower-mass
mini-halos would likely have had a truncated star-formation history,
relative to the higher-mass mini-halos, since they would not have been
able to retain gas once star formation commenced. Although further
investigation is required, this may account in a natural way for the
apparent lack of Galactic GCs with metallicity below [Fe/H] $\sim
-$2.3. The higher-mass primordial sub-Galactic fragments could be
associated with small metal-poor galaxies at high redshift, such as
the Lyman-$\alpha$ emitting galaxies (LAE) described by Elmegreen et
al. (2012). Since it is expected that CN-strong stars {\it require} the
dense environment of GCs in order to form, their observed properties
strongly suggest that a significant fraction of GCs have been stripped
or disrupted in the IHR. In this context, the similarity of the global
properties of the metal-poor GCs, including their spatial, kinematics,
and orbital properties, to those of the CN-strong stars is especially
intriguing. Our present data certainly suggest a strong relationship
between these two samples -- both appear to be associated with the IHP
of the Milky Way.

\acknowledgments

TCB acknowledges partial support from grants PHY 02-16783 and PHY
08-22648: Physics Frontier Center / Joint Institute for Nuclear
Astrophysics (JINA), awarded by the U.S. National Science Foundation.
KF acknowledges Sungeun Kim for her advice about the
30 Dor nebula.

Funding for SDSS and SDSS-II has been provided by the Alfred P. Sloan
Foundation, the Participating Institutions, the National Science
Foundation, the U.S. Department of Energy, the National Aeronautics and
Space Administration, the Japanese Monbukagakusho, the Max Planck
Society, and the Higher Education Funding Council for England. The SDSS
Web Site is http://www.sdss.org/.

The SDSS is managed by the Astrophysical Research Consortium for the
Participating Institutions. The Participating Institutions are the
American Museum of Natural History, Astrophysical Institute Potsdam,
University of Basel, University of Cambridge, CaseWestern Reserve
University, University of Chicago, Drexel University, Fermilab, the
Institute for Advanced Study, the Japan Participation Group, Johns
Hopkins University, the Joint Institute for Nuclear Astrophysics, the
Kavli Institute for Particle Astrophysics and Cosmology, the Korean
Scientist Group, the Chinese Academy of Sciences (LAMOST), Los Alamos
National Laboratory, the Max-Planck-Institute for Astronomy (MPIA), the
Max-Planck-Institute for Astrophysics (MPA), New Mexico State
University, Ohio State University, University of Pittsburgh, University
of Portsmouth, Princeton University, the United States Naval
Observatory, and the University of Washington.

%% To help institutions obtain information on the effectiveness of their
%% telescopes, the AAS Journals has created a group of keywords for telescope
%% facilities. A common set of keywords will make these types of searches
%% significantly easier and more accurate. In addition, they will also be
%% useful in linking papers together which utilize the same telescopes
%% within the framework of the National Virtual Observatory.
%% See the AASTeX Web site at http://www.journals.uchicago.edu/AAS/AASTeX
%% for information on obtaining the facility keywords.

%% After the acknowledgments section, use the following syntax and the
%% \facility{} macro to list the keywords of facilities used in the research
%% for the paper.  Each keyword will be checked against the master list during
%% copy editing.  Individual instruments or configurations can be provided
%% in parentheses, after the keyword, but they will not be verified.

%% Appendix material should be preceded with a single \appendix command.
%% There should be a \section command for each appendix. Mark appendix
%% subsections with the same markup you use in the main body of the paper.

%% Each Appendix (indicated with \section) will be lettered A, B, C, etc.
%% The equation counter will reset when it encounters the \appendix
%% command and will number appendix equations (A1), (A2), etc.

{\it Facilities:} \facility{SDSS}.

\clearpage

\begin{deluxetable*}{ccrrrrrrrrrrr}
\tablewidth{0pt}
\tabletypesize{\scriptsize}
\tablenum{1}
\tablecolumns{13}
\tablecaption{Parameters for the Martell et al. (2011) Sample}
\tablehead{
\colhead{Name} &
\colhead{} &
\colhead{d} &
\colhead{r} &
\colhead{PM$_{RA}$} &
\colhead{e$_{PM_{RA}}$} &
\colhead{PM$_{DE}$} &
\colhead{e$_{PM_{DE}}$} &
\colhead{V$_{RAD}$} &
\colhead{eV$_{RAD}$} &
\colhead{[Fe/H]} &
\colhead{$\delta S(3839)$} &
\colhead{CN} \\
\colhead{} &
\colhead{} &
\colhead{(kpc)} &
\colhead{(kpc)} &
\colhead{(mas~yr$^{-1}$)} &
\colhead{(mas~yr$^{-1}$)} &
\colhead{(mas~yr$^{-1}$)} &
\colhead{(mas~yr$^{-1}$)} &
\colhead{(km~s$^{-1}$)} &
\colhead{(km~s$^{-1}$)} &
\colhead{} &
\colhead{} &
\colhead{}\\
}
\startdata
   J025155.2-005526.4 &  & $17.1$ & $23.4$ & $2.23$ & $2.80$ & $-3.88$ & $2.80$ & $
-54.5$ & $1.1$  & $-1.32$ & $-0.03$ &        0 \\
J025141.7-003802.4  &  &$13.8$ & $20.3$ & $3.98$ & $2.72$ & $-3.98$ & $2.72$ & $
-150.6$ & $1.2$  & $-0.97$ & $0.09$ &        0 \\
J024814.1-004106.0  &  &$42.4$ & $48.1$ & $-0.72$ & $2.84$ & $-3.64$ & $2.84$ & $
23.8$ & $1.9$  & $-1.75$ & $-0.01$ &        0 \\
J024954.9-001716.8 &  & $27.2$ & $33.2$ & $1.38$ & $2.65$ & $-3.43$ & $2.65$ & $
-166.0$ & $1.4$  & $-1.93$ & $0.00$ &        0 \\
J024958.0-000003.6 &  & $40.9$ & $46.7$ & $0.61$ & $3.31$ & $-0.58$ & $3.31$ & $
-120.9$ & $2.3$  & $-1.55$ & $-0.03$ &        0 \\
J024959.7-001525.2 &  & $16.5$ & $22.8$ & $6.90$ & $2.73$ & $-3.14$ & $2.73$ & $
39.8$ & $1.7$  & $-1.17$ & $0.00$ &        0 \\
J024850.8-002830.0 &  & $28.1$ & $34.1$ & $-0.66$ & $3.19$ & $-3.77$ & $3.19$ & $
-130.5$ & $2.$  &$-1.23$ & $0.00$ &        0 \\
J024625.2-005436.0 &  &$7.8$ & $14.6$ & $5.15$ & $2.74$ & $-4.70$ & $2.74$ & $
-82.6$ & $0.8$  &$-1.19$ & $0.02$ &        0 \\
J024536.2-000636.0 &  &$20.4$ & $26.5$ & $-4.22$ & $3.01$ & $-2.82$ & $3.01$ & $
-152.9$ & $1.5$  &$-1.23$ & $-0.06$ &        0 \\
J024511.0-002031.2 &  &$5.8$ & $12.9$ & $9.05$ & $2.61$ & $-3.99$ & $2.61$ & $
272.9$ & $1.0$  &$-1.76$ & $0.02$ &        0 \\
                 &   &      &      &     &      &         &         &         &         &         &         &           \\
\enddata
\tablecomments{The full table is available in electronic form only.
The last column indicates whether the star is considered CN normal (0),
or CN strong (1).}
\end{deluxetable*}

\begin{deluxetable*}{crrrrrrrrrrr}
\tablewidth{0pt}
\tabletypesize{\scriptsize}
\tablenum{2}
\tablecolumns{12}
\tablecaption{Kinematic and Orbital Parameters for the Martell et al. (2011) Sample}
\tablehead{
\colhead{Name} &
\colhead{U} &
\colhead{e$_{U}$} &
\colhead{V} &
\colhead{e$_{V}$} &
\colhead{W} &
\colhead{e$_{W}$} &
\colhead{V$_{\phi}$} &
\colhead{e$_{V_{\phi}}$} &
\colhead{r$_{max}$} &
\colhead{r$_{min}$} &
\colhead{Z$_{max}$} \\
\colhead{} &
\colhead{(km~s$^{-1}$)} &
\colhead{(km~s$^{-1}$)} &
\colhead{(km~s$^{-1}$)} &
\colhead{(km~s$^{-1}$)} &
\colhead{(km~s$^{-1}$)} &
\colhead{(km~s$^{-1}$)} &
\colhead{(km~s$^{-1}$)} &
\colhead{(km~s$^{-1}$)} &
\colhead{(kpc)} &
\colhead{(kpc)} &
\colhead{(kpc)} \\
}
\startdata
J025155.2-005526.4 & $-89.9$ & $176.8$ & $-345.1$ & $237.5$ & $-9.4$ & $143.3
$ & $-121.3$ & $237.4$ & $10.0$ & $25.4$ & $16.9$ \\
J025141.7-003802.4 & $-74.9$ & $138.9$ & $-363.7$ & $193.0$ & $123.2$ & $112.8
$ & $-140.5$ & $193.1$ & $9.4$ & $27.3$ & $16.8$ \\
J024814.1-004106.0 & $-442.7$ & $457.8$ & $-438.5$ & $578.3$ & $-401.6$ & $364.8
$ & $-185.7$ & $578.0$ & \nodata & \nodata & \nodata \\
J024954.9-001716.8 & $-220.7$ & $266.8$ & $-445.6$ & $352.6$ & $14.1$ & $216.9
$ & $-211.0$ & $352.5$ & $22.5$ & $84.2$ & $63.8$ \\
J024958.0-000003.6 & $-68.7$ & $498.8$ & $-158.8$ & $642.2$ & $100.5$ & $407.2
$ & $66.2$ & $643.4$ & $8.1$ & $54.2$ & $40.0$ \\
J024959.7-001525.2 & $224.1$ & $171.8$ & $-528.0$ & $239.9$ & $102.4$ & $137.6
$ & $-319.4$ & $241.5$ & $21.3$ & $107.2$ & $75.8$ \\
J024850.8-002830.0 & $-390.2$ & $338.0$ & $-313.2$ & $429.4$ & $-155.9$ & $271.7
$ & $-67.4$ & $429.3$ & \nodata & \nodata & \nodata \\
J024625.2-005436.0 & $-28.1$ & $80.5$ & $-249.6$ & $113.6$ & $76.3$ & $62.6
$ & $-28.6$ & $113.7$ & $1.4$ & $15.1$ & $9.8$ \\
J024536.2-000636.0 & $-487.8$ & $241.5$ & $65.7$ & $290.7$ & $-170.5$ & $190.7
$ & $320.2$ & $291.8$ & \nodata & \nodata & \nodata \\
J024511.0-002031.2 & $259.4$ & $60.1$ & $-214.1$ & $86.9$ & $-151.4$ & $45.9
$ & $-3.2$ & $88.1$ & $0.1$ & $33.0$ & $16.2$ \\
                 &      &      &     &      &         &         &         &         &         &         &             \\
\enddata
\tablecomments{The full table is available in electronic form only}
\end{deluxetable*}

\begin{deluxetable*}{crrrrrrrrrrr}
\tablewidth{0pt}
\tabletypesize{\scriptsize}
\tablenum{3}
\tablecolumns{12}
\tablecaption{Kinematic and Orbital Parameters for the Sample of Globular Clusters with Available Proper Motions}
\tablehead{
\colhead{Name} &
\colhead{U} &
\colhead{e$_{U}$} &
\colhead{V} &
\colhead{e$_{V}$} &
\colhead{W} &
\colhead{e$_{W}$} &
\colhead{V$_{\phi}$} &
\colhead{e$_{V_{\phi}}$} &
\colhead{r$_{max}$} &
\colhead{r$_{min}$} &
\colhead{Z$_{max}$} \\
\colhead{(km~s$^{-1}$)} &
\colhead{(km~s$^{-1}$)} &
\colhead{(km~s$^{-1}$)} &
\colhead{(km~s$^{-1}$)} &
\colhead{(km~s$^{-1}$)} &
\colhead{(km~s$^{-1}$)} &
\colhead{(km~s$^{-1}$)} &
\colhead{(km~s$^{-1}$)} &
\colhead{(kpc)} &
\colhead{(kpc)} &
\colhead{(kpc)} &
\colhead{} \\
}
\startdata
              NGC~104    &    84.6    &    10.5    &   $-$75.0    &    11.7    &    47.8    &     5.3    &   166.0    &     6.9    &     7.9    &     6.0    &     3.6\\
               NGC~288    &    26.9    &    11.8    &  $-$293.9    &    33.7    &    51.2    &     0.3    &   $-$74.0    &    33.7    &    12.1    &     3.6    &     9.9\\
               NGC~362    &    25.9    &    25.6    &  $-$296.0    &    29.3    &   $-$68.6    &    24.8    &   $-$37.5    &    24.4    &    10.6    &     1.2    &     7.0\\
              NGC~3201    &  $-$183.0    &    13.7    &  $-$450.7    &     2.7    &   130.4    &     9.0    &  $-$292.2    &     6.7    &    21.9    &     9.2    &     7.4\\
              NGC~4372    &   107.9    &    17.7    &  -158.0    &    12.3    &    76.8    &    12.1    &   117.8    &     5.7    &     7.6    &     3.4    &     2.3\\
              NGC~4833    &    95.0    &    24.4    &  $-$291.4    &    15.9    &   $-$42.6    &    10.8    &    23.0    &    18.1    &     8.6    &     0.5    &     1.3\\
              NGC~5024    &   $-$38.3    &    83.9    &    34.7    &    84.6    &   $-$59.8    &    15.1    &   237.3    &    84.3    &    31.5    &    15.8    &    30.2\\
               NGC~5139    &   $-$55.9    &    11.4    &  $-$261.9    &    11.1    &    $-$0.1    &    10.3    &   $-$66.8    &     4.4    &     6.7    &     1.5    &     1.3\\
              NGC~5272    &   $-$64.2    &    26.6    &  $-$111.3    &    25.7    &  $-$135.1    &     5.7    &   118.8    &    26.4    &    16.1    &     5.1    &    13.7\\
                     Pal 5&    78.5    &    14.2    &  $-$333.6    &    39.0    &    15.1    &    13.9    &   111.0   &    39.1    &    18.1    &     9.1    &    17.4\\
               NGC~5897    &    33.5    &    31.5    &  $-$304.9    &    57.3    &   118.6    &    43.8    &    71.3    &    56.3    &     8.8    &     1.9    &     8.1\\
              NGC~5904    &  $-$329.3    &    41.6    &  $-$184.3    &    46.2    &  $-$207.7    &    36.8    &    69.0    &    39.9    &    37.5    &     1.0    &    32.0\\
              NGC~5927    &   207.6    &    15.4    &  $-$116.0    &    22.8    &    36.8    &    15.0 &   232.2    &     3.8    &     6.3    &     4.7    &     0.9\\
              NGC~5986    &     5.4    &    13.7    &  $-$241.2    &    28.3    &    31.5    &    19.4    &     9.6    &    21.6    &     4.5    &     0.2    &     2.4\\
               NGC~6093    &    $-$8.8    &    10.1    &  $-$348.6    &    46.7    &   $-$98.4    &    29.9    &    67.1    &    76.0    &     3.6    &     1.6    &     4.9\\
              NGC~6121    &   $-$49.2    &     2.6    &  $-$235.7    &    21.7    &   $-$10.5    &     4.4    &   $-$18.3    &    21.9    &     6.6    &     0.4    &     0.6\\
               NGC~6144    &  $-$171.5    &     8.5    &  $-$260.4    &    38.0    &    14.4    &    27.8    &  $-$172.1    &    30.2    &     4.1    &     1.4    &     3.2\\
              NGC~6171    &     1.5    &    11.6    &   $-$76.4    &    30.2    &   $-$43.3    &    26.4    &   142.1    &    29.9    &     3.4    &     3.0    &     2.7\\
              NGC~6205    &   227.1    &    28.7    &   $-$77.3    &    21.7    &  $-$137.2    &    20.3    &   $-$30.7    &    28.6    &    21.1    &     5.4    &    21.3\\
              NGC~6218    &   $-$59.2    &    21.8    &  $-$115.4    &    41.3    &  $-$114.4    &    39.8    &   116.3    &    37.7    &     5.4    &     2.9    &     3.3\\
              NGC~6254    &   $-$95.5    &    13.2    &  $-$117.3    &    28.4    &    82.1    &    19.8    &   121.5    &    24.3    &     5.3    &     3.0    &     2.7\\
              NGC~6266    &    79.4    &     3.1    &   $-$68.0    &    14.9    &    71.2    &    14.0    &   170.9    &    11.6    &     3.3    &     1.9    &     1.2\\
              NGC~6273    &  $-$123.5    &     5.7    &   $-$80.7    &    22.7    &   112.4    &    22.2    &  -163.2    &   141.4    &     4.2    &     0.5    &     2.6\\
              NGC~6284    &   $-$25.3    &     8.9    &  -460.9    &    72.5    &    $-$4.7    &    50.9    &   238.7    &    72.6    &    10.2    &     6.7    &     3.6\\
              NGC~6287    &   281.6    &     7.8    &  $-$212.5    &    40.1    &    $-$8.3    &    36.1    &   $-$15.6    &    45.6    &     8.0    &     0.1    &     3.9\\
              NGC~6293    &   123.5    &     5.8    &  $-$163.8    &    39.2    &  $-$156.1    &    38.9    &    $-$2.5    &    45.4    &     3.5    &     0.0    &     2.6\\
               NGC~6304    &   106.3    &     3.8    &   $-$57.3    &    11.3    &    30.6    &     8.8    &   177.4    &     8.7    &     4.1    &     2.2    &     0.7\\
               NGC~6316    &   $-$69.0    &     9.5    &  $-$166.6    &    33.3    &    36.8    &    29.9    &   $-$69.9    &    36.1    &     2.3    &     1.0    &     1.1\\
              NGC~6333    &  $-$256.2    &     8.3    &   $-$91.9    &    23.3    &    $-$8.2    &    21.1    &   270.3    &    47.4    &     7.7    &     1.1    &     3.1\\
              NGC~6341    &    20.2    &    22.7    &  $-$166.7    &    22.9    &    51.0    &    30.7    &    21.9    &    21.8    &    10.3    &     0.7    &     5.2\\
              NGC~6342    &  $-$150.8    &     6.0    &  $-$234.6    &    38.2    &    $-$9.7    &    27.7    &   144.4    &    52.7    &     2.8    &     0.7    &     1.7\\
              NGC~6356    &   $-$67.2    &     9.9    &  $-$325.0    &    53.4    &    57.3    &    45.2    &   119.2    &    51.8    &     7.3    &     3.3    &     3.1\\
               NGC~6362    &    92.1    &    14.0    &  $-$130.2    &    21.0    &    41.1    &    16.1    &   125.5    &     6.8    &     5.3    &     3.0    &     2.4\\
              NGC~6388    &   $-$34.0    &     8.4    &  $-$200.7    &    29.6    &   $-$21.7    &    21.9    &   $-$38.8    &    18.0    &     2.7    &     0.7    &     1.2\\
              NGC~6397    &    41.3    &     6.0    &   $-$99.4    &     9.7    &  $-$108.7    &    10.3    &   124.9   &     8.5    &     6.6    &     3.4    &     2.2\\
              NGC~6441    &    $-$1.3    &     6.5    &  $-$232.0    &    45.3    &    44.7    &    30.2    &    10.4    &    44.3    &     3.3    &     0.2    &     1.1\\
               NGC~6584    &   $-$70.5    &    26.6    &  $-$371.1    &    52.2    &  $-$184.3    &    41.0    &    54.1    &    64.5    &    13.4    &     1.2    &     9.0\\
              NGC~6626    &   $-$40.1    &     3.4    &  $-$173.6    &    26.8    &  $-$110.0    &    21.4    &    54.5    &    24.0    &     3.2    &     1.0    &     1.2\\
              NGC~6656    &   153.4    &     2.3    &    $-$47.2    &     6.1    &  $-$111.4    &    14.4    &   196.2    &     6.3    &    11.2    &     3.3    &     2.4\\
              NGC~6712    &    99.8    &     5.8    &   $-$35.6    &    11.8    &  $-$136.6    &    20.2    &    34.0    &    39.0    &     7.4    &     0.4    &     2.1\\
              NGC~6723    &    88.8    &     6.6    &   $-$73.6    &    22.2    &    10.4    &    18.2    &   141.5    &    37.6    &     4.0    &     0.2    &     3.6\\
               NGC~6752    &    34.0    &     5.1    &   -29.8    &     9.4    &    21.3    &     7.2    &   192.3    &     8.6    &     6.0    &     5.1    &     1.8\\
              NGC~6779    &   108.5    &    40.1    &   $-$80.9    &    21.4    &     3.1    &    44.1    &   $-$33.2    &    32.6    &    12.9    &     0.8    &     1.9\\
              NGC~6809    &  $-$207.9    &    11.1    &  $-$222.6    &    32.4    &   $-$61.4    &    21.4    &    40.3    &    21.4    &     7.3    &     0.7    &     4.5\\
              NGC~6838    &   $-$84.2    &    15.4    &   $-$65.6    &    10.2    &    $-$3.0    &    15.2    &   175.8    &     2.1    &     7.1    &     4.9    &     0.3\\
              NGC~6934    &    69.7    &    62.4    &  $-$533.0    &    54.9    &  $-$120.2    &    74.9    &   $-$54.4    &    74.5    &    40.9    &     6.4    &    39.8\\
              NGC~7006    &  $-$111.8    &    71.4    &  $-$437.2    &    43.4    &   148.1    &    68.0    &   160.9    &    79.5    &    82.9    &    16.9    &    42.7\\
              NGC~7078    &  $-$227.8    &    60.3    &  $-$291.3    &    55.6    &  $-$113.1    &    56.0    &   165.4    &    37.2    &    18.4    &     8.3    &    13.9\\
              NGC~7089    &   100.8    &    42.0    &  $-$215.5    &    42.2    &  $-$328.4    &    51.5    &   $-$92.2    &    40.8    &    35.5    &     6.4    &    34.0\\
               NGC~7099    &    65.2   &    21.2    &  $-$330.2    &    36.8    &    51.0    &    20.1    &  $-$127.6    &    28.2    &     7.1    &     4.2    &     6.2\\
                    %Pal 12&  -218.560    &    29.250    &  -322.010    %&    42.230    &   -20.600    &    18.190    &   %240.660    &    40.350    &    21.216    &    %15.221    &    19.237\\
                    Pal 13&   251.4    &    41.1    &   $-$40.0    &    22.2    &  $-$100.7    &    24.6    &  $-$167.7    &    47.8    &    77.4    &    12.1    &    57.0\\
                 &      &      &     &      &         &         &         &         &         &         &          \\
\enddata
%\tablecomments{The errors for the distances in the SDSS-SEGUE data is $~$ 10-15\%}
\end{deluxetable*}

\clearpage

\newpage
\begin{figure}[ht]
\figurenum{1}
%\hspace{1cm}
\begin{minipage}[b]{0.48\linewidth}
\centering
\includegraphics[width=\textwidth]{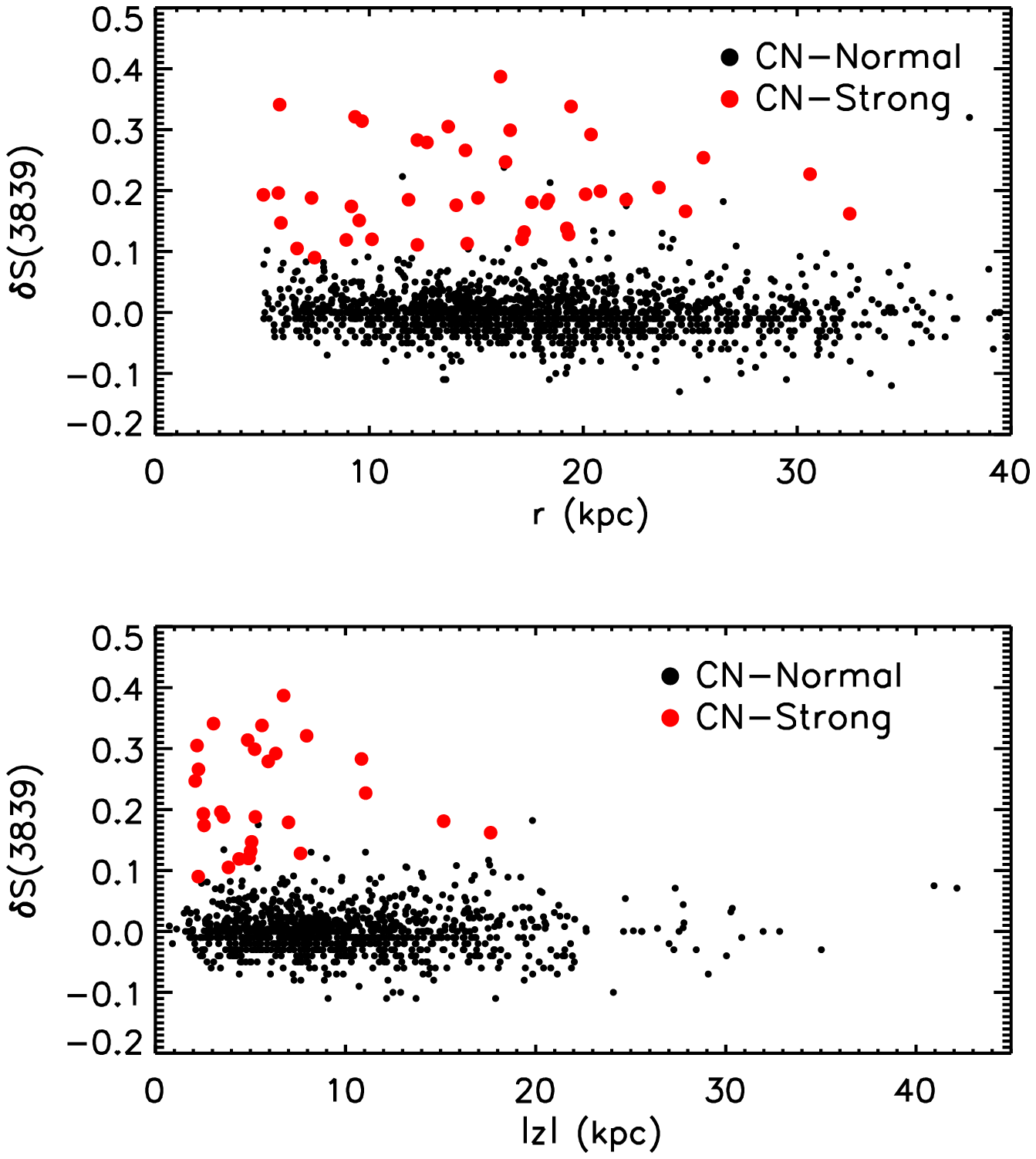}
%\label{fig:figure1}
\end{minipage}
%\hspace{5cm}
\begin{minipage}[b]{0.48\linewidth}
\centering
\includegraphics[width=\textwidth]{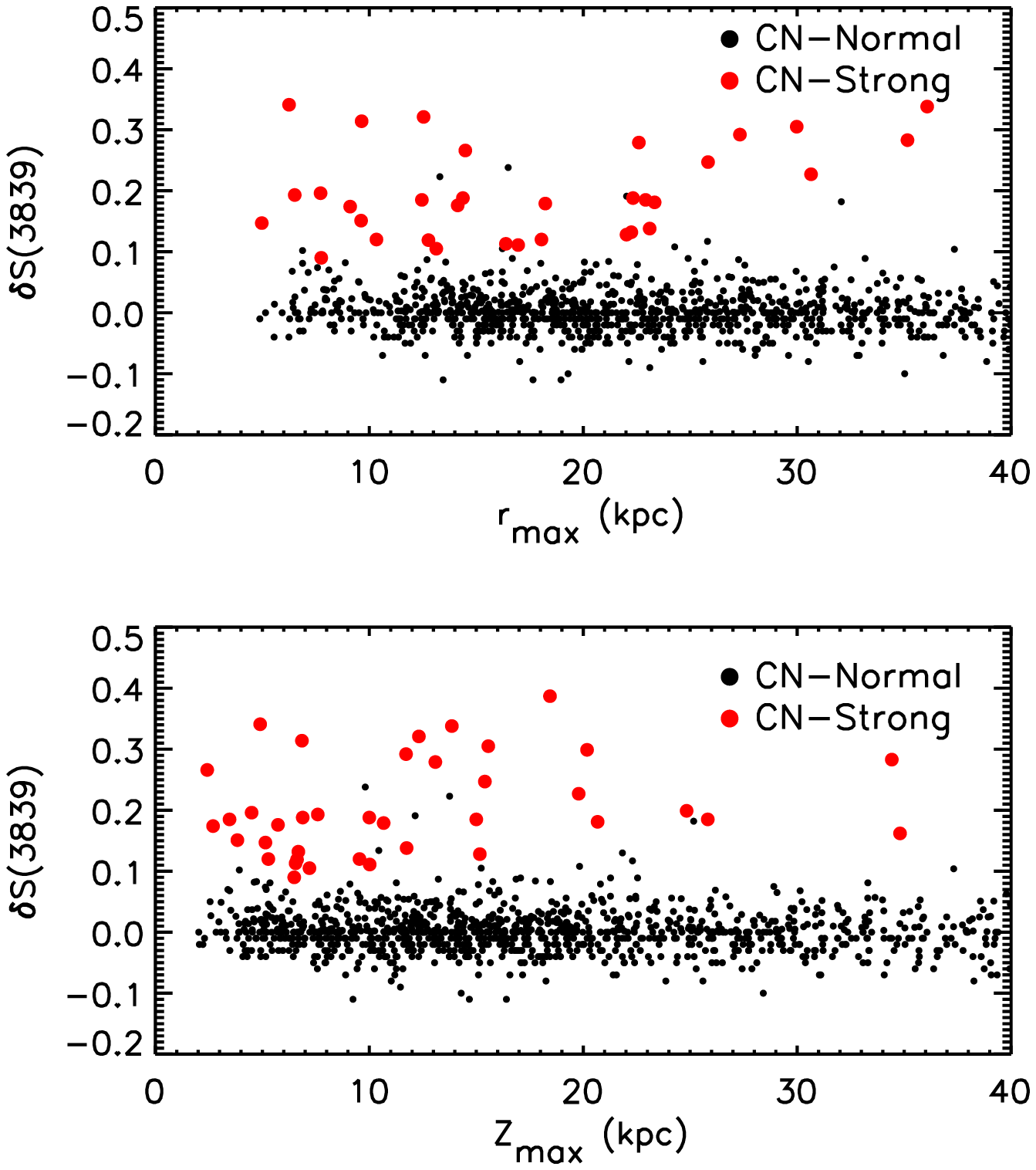}
%\caption{Metallicity as a function of
%the Galactocentric radius, $r$, for the 59 globular clusters with
%available proper motions.}
%\label{fig:figure2}
\end{minipage}
\caption{Top-left panel: $\delta$S(3839) index as a function of the Galactocentric distance, $r$.
Black dots represent the CN-normal stars selected from the Martell et
al. (2011) sample, while the red dots indicate CN-strong stars. Bottom-left
panel: $\delta$S(3839) index as a function of the vertical distance,
$|z|$. Top-right panel: $\delta$S(3839) index, as a function of
apogalactic distance, $r_{max}$, for the CN-normal stars (black dots) and
CN-strong stars (red dots). Bottom-right panel: $\delta$S(3839) index, as
a function of the distance $Z_{max}$ (the maximum distance of a stellar
orbit above or below the Galactic plane), for the CN-normal stars (black
dots) and the CN-strong stars (red dots).}
\end{figure}

\begin{figure}
\centering
\figurenum{2}
\includegraphics[width=0.6\textwidth]{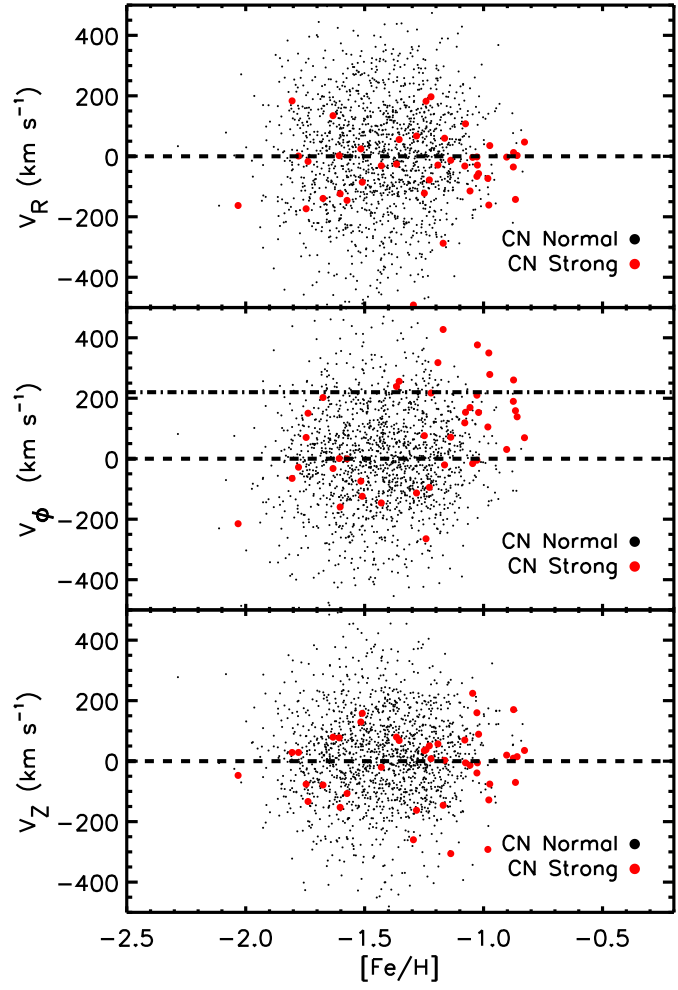}
\caption{Distribution of the velocity components (V$_{R}$,V$_{\phi}$,V$_{Z}$) versus [Fe/H]
for the stars selected from the Martell et al. (2011) sample. The dot-dashed
line in the middle panel is the adopted LSR velocity for stars in the
Solar Neighborhood, while the dashed line in each panel represents the mean velocity of a non-rotating population. Black dots
represent the CN-normal stars, while the red dots indicate the CN-strong stars.}
\end{figure}

\newpage

\begin{figure}
\centering
\figurenum{3}
\includegraphics[width=0.8\textwidth]{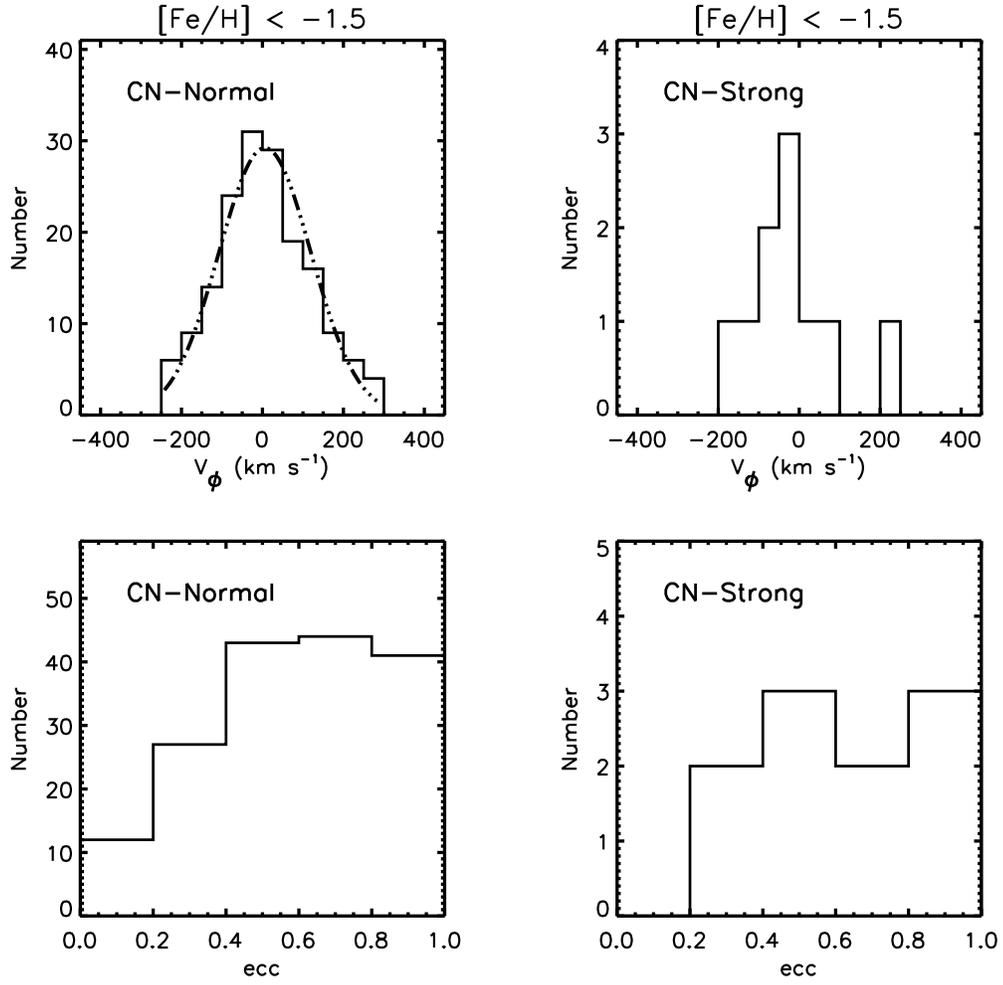}
%\end{figure}
%\begin{figure}
%\centering
%\figurenum{4}
%\includegraphics[angle=90,width=0.7\textwidth]{ecc_feha1.ps}
\caption{Top-left panel: Rotational velocity distribution in the Galactocentric
cylindrical reference frame, V$_{\phi}$, for the CN-normal stars at low
metallicity, [Fe/H] $< -$1.5. The dot-dashed curve indicates a Gaussian
fit to the distribution. Top-right panel: V$_{\phi}$ distribution for
the low-metallicity CN-strong stars. Bottom-left panel: Eccentricity
distribution for the CN-normal stars at low metallicity, [Fe/H] $<
-$1.5. Bottom-right panel: Eccentricity distribution for the
low-metallicity CN-strong stars.}
\end{figure}

\newpage
\begin{figure}[!ht]
\figurenum{4}
\hspace{3cm}
\vspace{1cm}
\begin{minipage}[t]{0.55\linewidth}
\centering
\includegraphics[angle=90,width=0.9\textwidth]{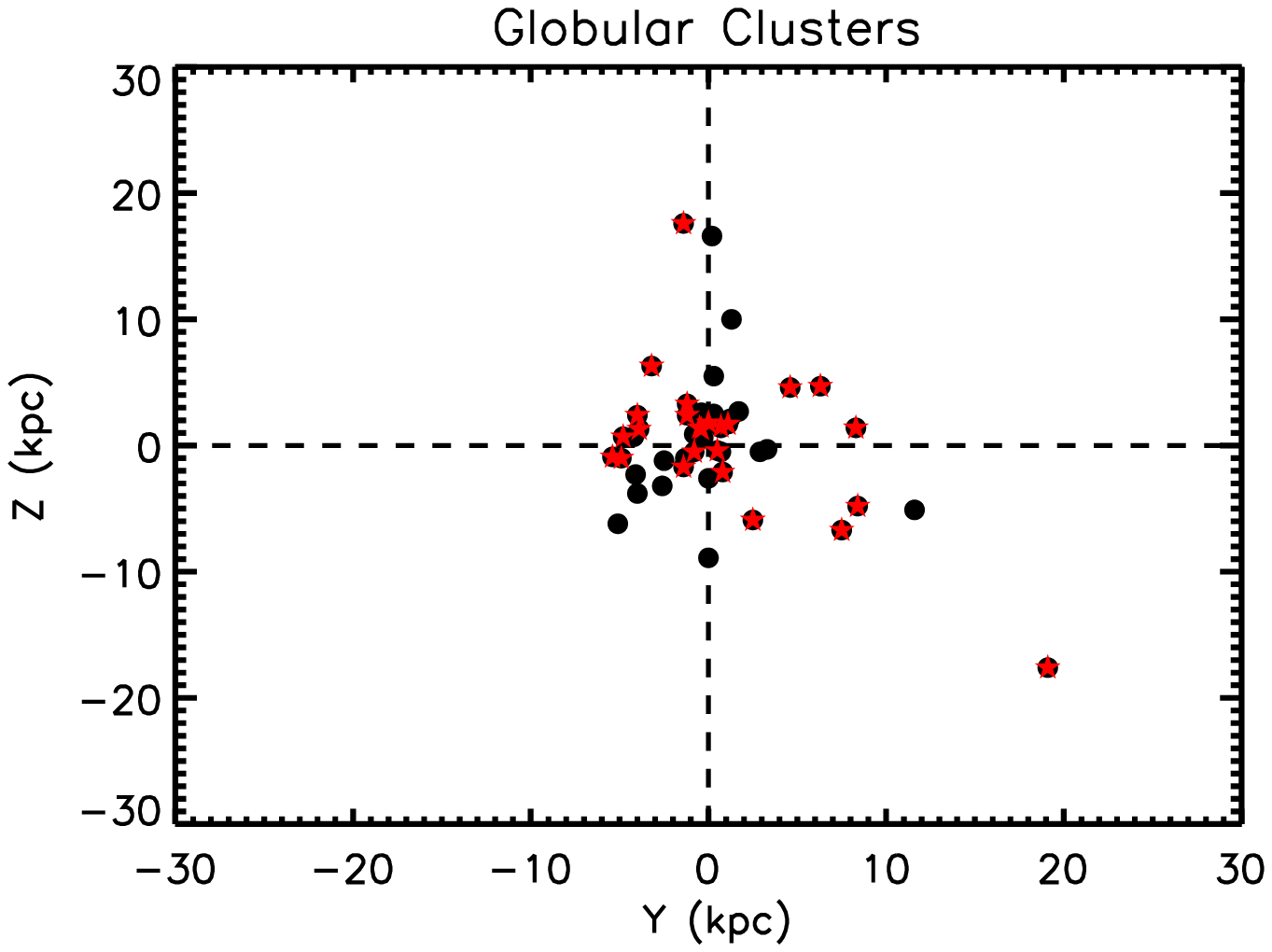}
%\label{fig:figure1}
\end{minipage}
\hspace{-6cm}
%\vspace{2cm}
\begin{minipage}[b]{0.6\linewidth}
%\centering
\hspace{1cm}
\vspace{0.55cm}
\includegraphics[width=0.9\textwidth]{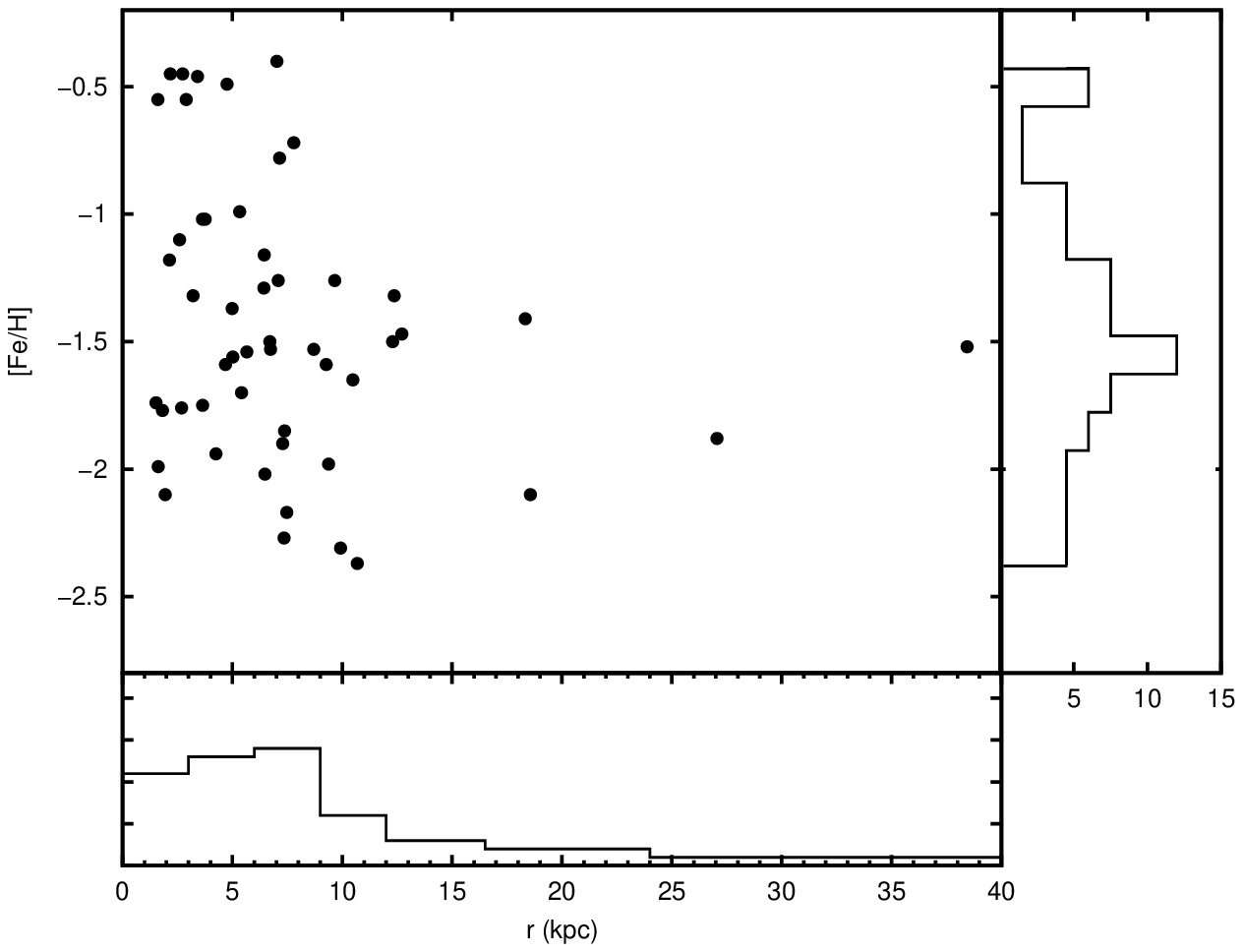}
%\caption{Metallicity as a function of
%the Galactocentric radius, $r$, for the 59 globular clusters with
%available proper motions.}
%\label{fig:figure2}
\end{minipage}
\hspace{3.5cm}
\vspace{6cm}
\begin{minipage}[b]{0.6\linewidth}
\hspace{3.5cm}
%\centering
\includegraphics[angle=90,width=0.85\textwidth]{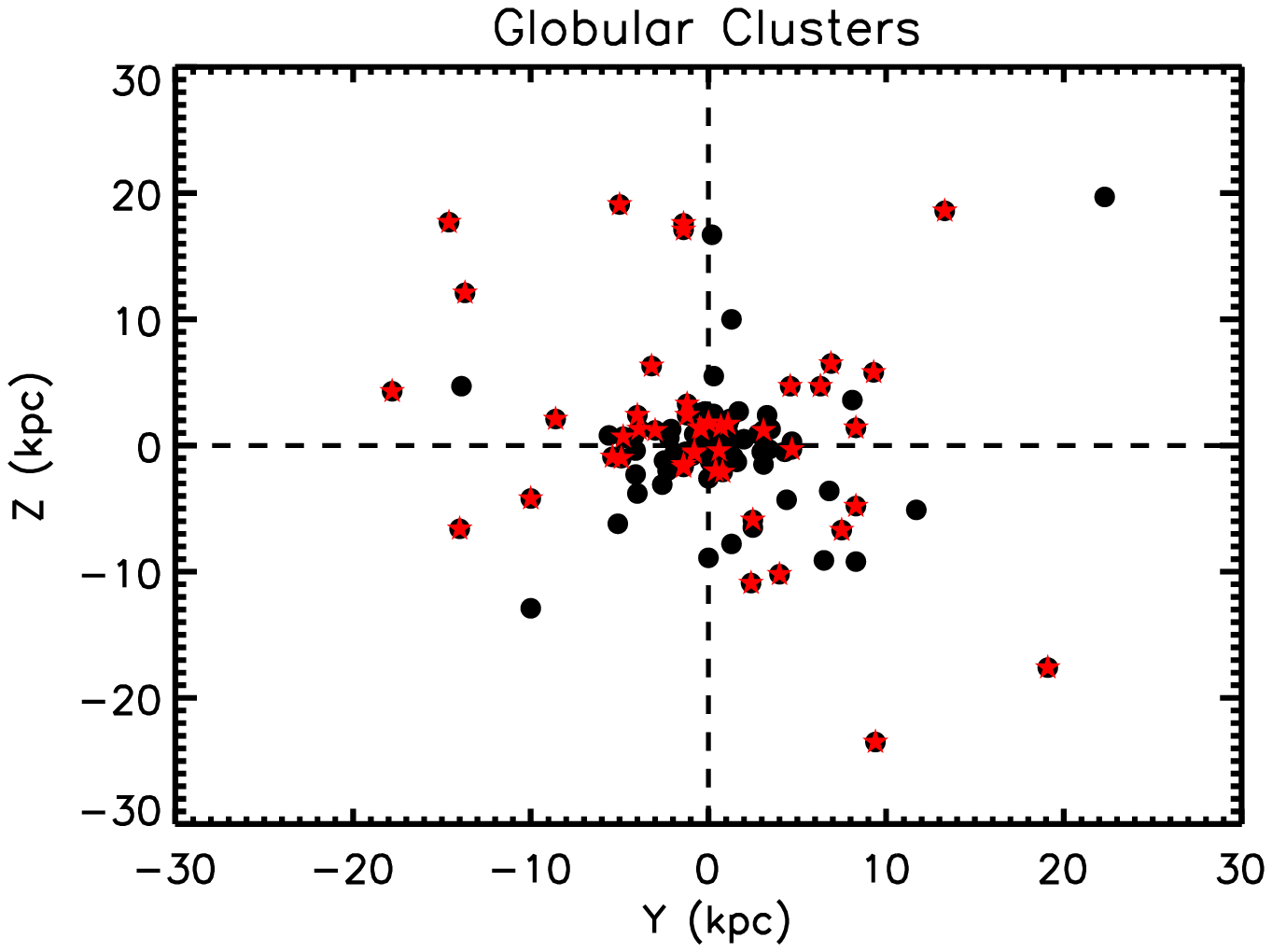}
%\caption{Metallicity as a function of
%the Galactocentric radius, $r$, for the 59 globular clusters with
%available proper motions.}
%\label{fig:figure2}
\end{minipage}
\hspace{-2.5cm}
\vspace{-6cm}
\begin{minipage}[!h]{0.6\linewidth}
%\centering
\vspace{-8.0cm}
\includegraphics[width=0.9\textwidth]{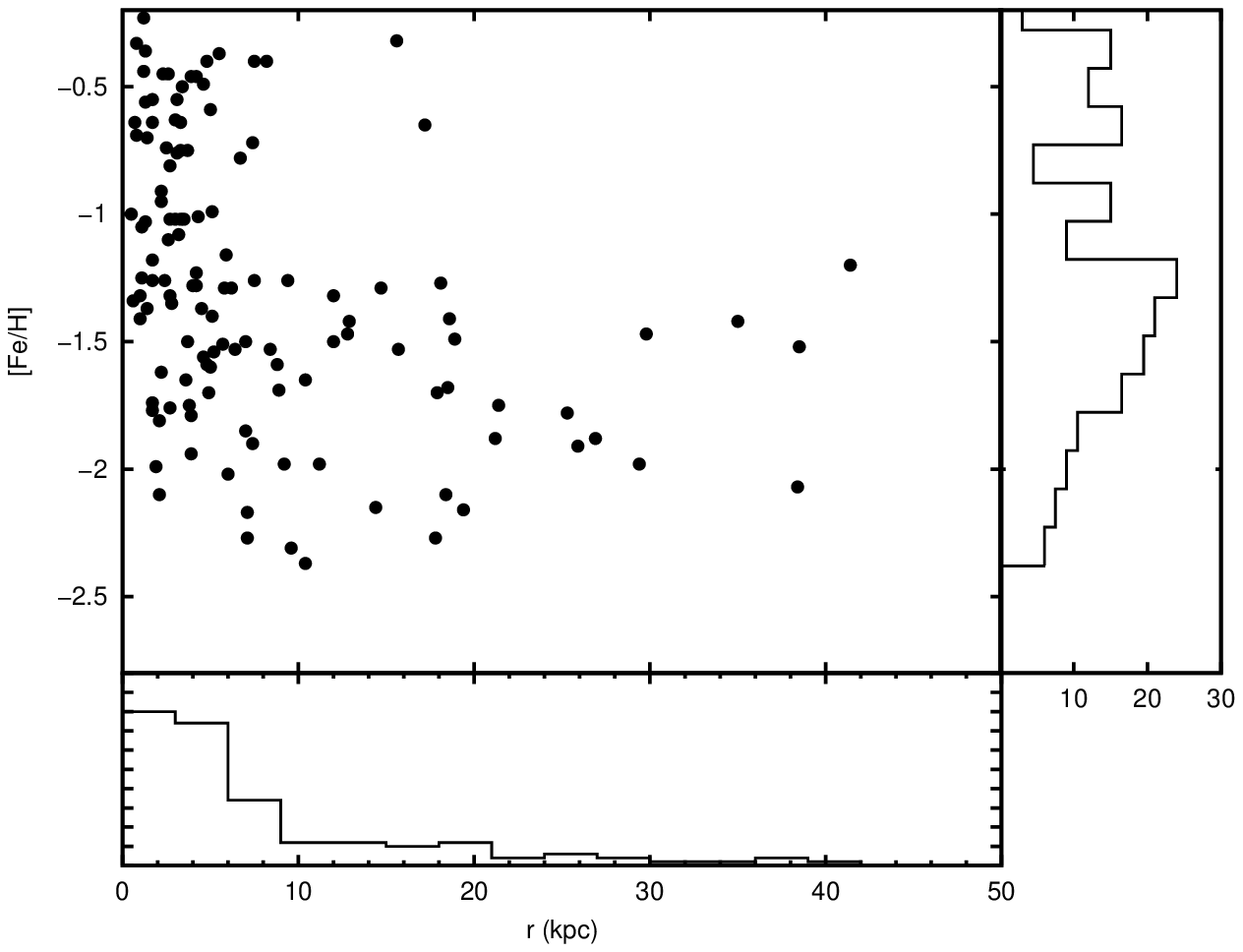}
%\caption{Metallicity as a function of
%the Galactocentric radius, $r$, for the 59 globular clusters with
%available proper motions.}
%\label{fig:figure2}
\end{minipage}
\caption{Top-left Panel: The distribution of the Galactic globular cluster sample with available
absolute proper motions, projected onto the YZ plane, in the Galactocentric Cartesian
reference system, with (0,0) at the Galactic center. Red dots indicate the globular clusters at low
metallicity, [Fe/H] $< -$1.5. Top-right Panel: Metallicity as a function of
the Galactocentric radius, $r$, for the 59 globular clusters with
available absolute proper motions. The marginal histograms denote the
distributions of [Fe/H] and $r$. Bottom-left Panel: The distribution of
the 129 Galactic globular clusters selected from the Harris (2006, 2010
update) database, projected onto the YZ plane. Bottom-right Panel: Metallicity as a
function of the Galactocentric radius, $r$, for the 129 globular
clusters.}
\end{figure}

\newpage

\begin{figure}
\centering
\figurenum{5}
\includegraphics[width=0.9\textwidth]{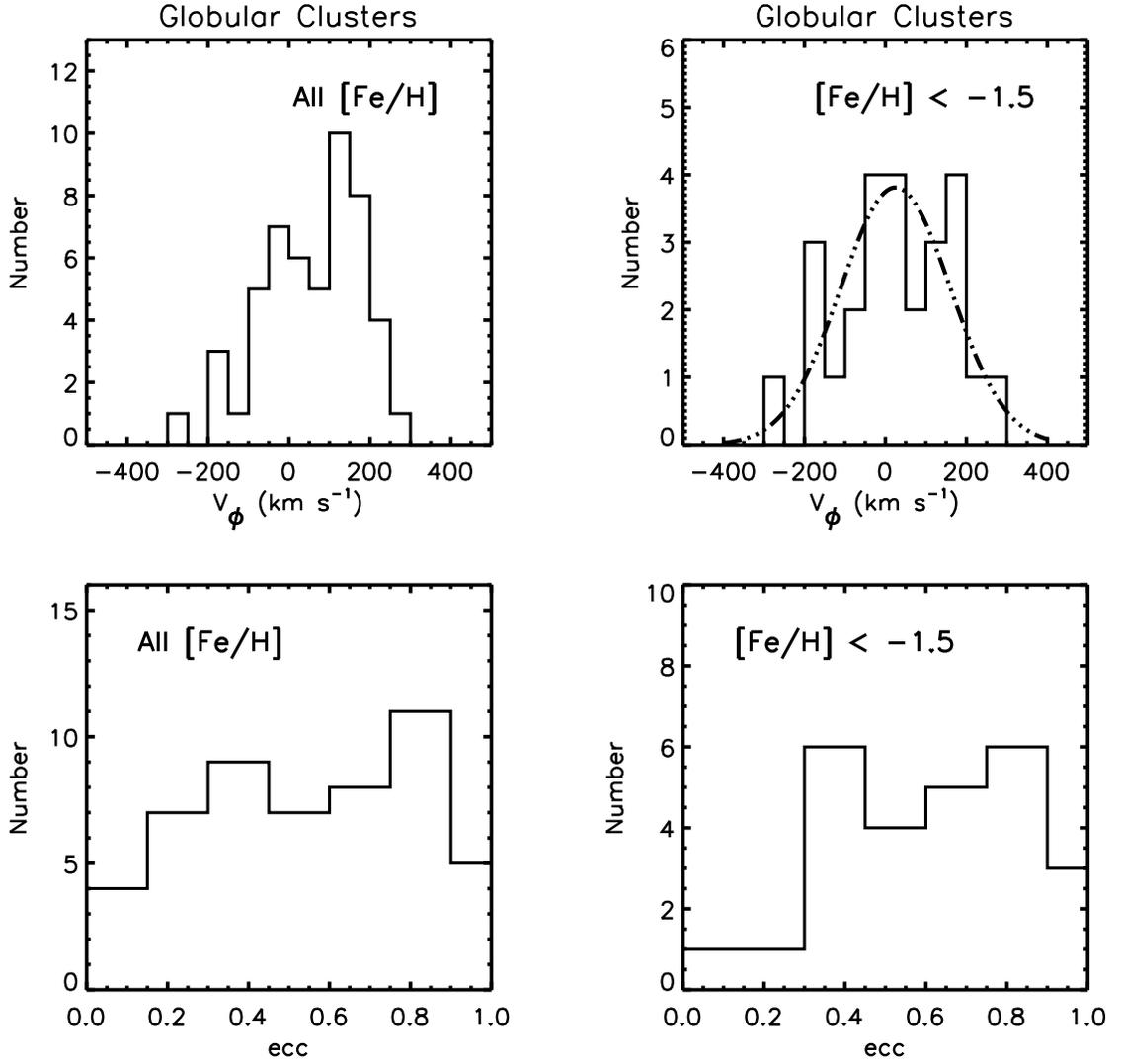}
\caption{Top-left panel: Rotational velocity distribution in the Galactocentric cylindrical
reference frame, V$_{\phi}$, for the sample of Galactic globular
clusters with available proper motions, and no selection on metallicity.
Top-right panel: The same as the left panel, but for the clusters at low
metallicity, [Fe/H] $< -$1.5. The dot-dashed curve indicates a Gaussian
fit to the distribution. Bottom-left panel: Eccentricity distribution
for the sample of globular clusters with available proper motions and no
selection on metallicity. Bottom-right panel: The same as in the left
panel, but for the clusters at low metallicity, [Fe/H] $< -$1.5.}
\end{figure}

\newpage

\end{document}